\documentclass[11pt,a4paper]{article}
\usepackage{amsmath}
\usepackage{enumerate}
\usepackage{amssymb}
\usepackage{graphicx}
\usepackage{multirow}
\usepackage{xcolor}
\usepackage{tablefootnote}
\usepackage{threeparttable}
\usepackage{bm}
\usepackage[normalem]{ulem}
\usepackage{physics}
\usepackage{cancel}

\usepackage{amsmath,mathtools}
\usepackage{float}

\usepackage{jcappub}

\begin{document}

\title{An Entangled Universe}
\author[a]{Pablo Tejerina-P\'erez,}
\author[b,c,d]{Daniele Bertacca,}
\author[a,e]{Raul Jimenez,}

\affiliation[a]{ICCUB, University of Barcelona, Mart\' i i Franqu\` es, 1, E08028
Barcelona, Spain}
\affiliation[b]{Dipartimento di Fisica e Astronomia Galileo Galilei, Universit\`a degli Studi di Padova, via Marzolo 8, I-35131, Padova, Italy}
\affiliation[c]{INFN, Sezione di Padova, via Marzolo 8, I-35131, Padova, Italy}
\affiliation[d]{INAF-Osservatorio Astronomico di Padova, Vicolo dell Osservatorio 5, I-35122 Padova, Italy}
\affiliation[e]{ICREA, Pg. Lluis Companys 23, Barcelona, 08010, Spain.}

\emailAdd{pablo.tejerina@icc.ub.edu; daniele.bertacca@unipd.it; raul.jimenez@icc.ub.edu}

\abstract{We propose a possible quantum signature of the early Universe that could lead to observational imprints of the quantum nature of the inflationary period. Graviton production in the presence of an inflaton scalar field results in entangled states in polarization. This is because of a non-trivial effect
due to the derivatives on two scalar fluctuations and it provides a fingerprint that
depends on the polarization of the graviton that Alice and/or Bob measured in their patch. At horizon crossing, interactions between the gravitons and inflatons perform the required Bell experiments leading to a definitive measure.  We hint how this signature could be measured in the high-order correlation function of galaxies, in particular on the halo bias and the intrinsic alignment.}

\maketitle

\section{Introduction}
Inflation is a period of quasi-exponential expansion of the early Universe, i.e. a de Sitter stage with slightly broken time-translational symmetry to allow for a graceful exit \cite{Mukhanov:1981xt, Guth:1982ec, Hawking:1982cz, Bardeen:1983qw, Mukhanov_1}:
\begin{equation}
\label{eq: deSitter scale factor}
    a(t)\sim a_i\,e^{H_\Lambda t}\,\,,
\end{equation}
where $a$ is the scale factor, with initial value $a_i$ and $H_\Lambda = (8\pi G\varepsilon_\Lambda/3)^{1/2}$, where $\varepsilon_\Lambda$ is the constant energy density of a perfect fluid with positive cosmological constant $\Lambda$, with equation of state $p_\Lambda = -\varepsilon_\Lambda$.

The presence of an inflationary stage previous to the Big Bang (or what used to be called ``Big Bang'' before the upcoming theory of inflation) solves in a natural manner the fine tuning and naturalness problems of the previous theory (horizon problem, flatness problem/fine tuning of the initial conditions, and others - for reference, see for example \cite{Mukhanov_1}) while keeping its successful predictions like Big Bang Nucleosynthesis. It also introduces a natural way to explain the inhomogeneities and anisotropies in the Universe that gave rise to the formation of the cosmological structures we observe today. Quantum fluctuations in the fields present during inflation (in its most minimal form, just the spacetime-metric field, and a scalar field called the \textit{inflaton}) get stretched by the exponential expansion to cosmological distances. Their ``size'' (physical wavelength) grows proportional to $a$, while their amplitude decays as $a^{-1}$. The curvature scale $H_\Lambda^{-1}$ (or Hubble horizon) remains almost constant during inflation. Thus, for any mode of fixed comoving wavenumber $k$, the physical wavelength $\lambda_{ph}=2\pi(a/k)$ of a quantum fluctuation generated inside the Hubble horizon will soon become larger than $H_\Lambda^{-1}$. Therefore, the corresponding mode leaves the horizon, and starts ``feeling'' the curvature of spacetime. At horizon crossing, its amplitude freezes and remains almost constant until the end of the inflationary stage. Once inflation is over, the radiation dominated era starts (decelerating Hubble expansion), during which the curvature scale $H_\text{rad}^{-1}$ starts growing at a rate $a/\Dot{a}$. At this point, the now large fluctuations reenter the Hubble horizon as density perturbations of cosmological scale, and as gravitational waves. These density perturbations work as classical gravitational seeds for the formation of large-scale structures. Those modes that left the horizon the latest (in the last $8$ e-folds of inflation approximately, see \cite{Kolv_Turner}, sec. 8.4) will be the first ones to reenter, and are the ones relevant in determining the structures in our observable scales.

In this context, there is a somewhat natural question that emerges: at what point does a quantum fluctuation generated during inflation stop being of quantum nature? The analysis that we apply to the Cosmic Microwave Background (CMB), which provides us with the earliest observational features of the Universe so far, and to large scale structure survey's data, is classical. But where did the quantum nature of the gravitational seeds go? It is still an open question how this classicalization of quantum fluctuations occurs. If any signal of the early nature of the quantum fluctuation remains, this should be imprinted in some current observable, maybe in the form of non-gaussianities in the CMB anisotropies, or in higher-order correlation functions of galaxy distributions, or some other observable. 

There has been previous work in the literature trying to search for quantum signals of the early Universe\cite{Campo:2003pa,Campo:2003gb,Campo:2005sv,Campo:2005qn,Campo:2007ar,Campo:2008ju,Campo:2008ij,Choudhury:2016cso,Choudhury:2016pfr,Kanno:2015ewa,Martin:2015qta,Kanno:2017teu,Martin:2017zxs,Martin:2018zbe,Martin:2018lin,Kanno_2019,kanno2020polarized,Danielson:2021egj,Colas:2022kfu,Prabhu:2022zcr,DaddiHammou:2022itk,Ning_2023}. Quantum discord (a measure of ``\textit{quantumness}'') of inflationary perturbations is calculated in \cite{J.Martin_V.Venin} and suggests some features to probe different levels of discordance in CMB descriptions. Possible observational signatures in the CMB of graviton exchange between tensor and scalar fluctuations are discussed in \cite{Bellomo_2018}. On the other hand, due to the environment, the potential decoherence of quantumness upon classicalization has been widely  discussed in literature, e.g., see \cite{Calzetta:1995ys, Lombardo:2005iz, Martineau:2006ki, Kiefer:2006je, Kiefer:2008ku, Nelson:2016kjm, Burgess:2014eoa, Martin:2018zbe, Boyanovsky:2015tba, DaddiHammou:2022itk, Burgess_decoherence, Sou:2022nsd}. 
For example, in \cite{Collins_2016}, they discuss how the CMB polarization components $\bf E$ and $\bf B$ are modified due to entanglement of scalar and tensor fluctuations. Regarding primordial gravitons, a possible source of decoherence should also include the nonlinear interaction between tensor modes \cite{Gong:2019yyz} and the scalar-tensor interaction \cite{DaddiHammou:2022itk, Burgess_decoherence, Sou:2022nsd}. Another possibility is to add extra fields to the simplest model of inflation such that one can construct Bell inequalities as done in \cite{Maldacena_2015}.

In this work, we suggest a mechanism by which entangled states are created during inflation, via the interaction of gravitons and inflatons. We describe plausible processes by which the quantum nature of the tensor fluctuations of the metric field (i.e. gravitons) during inflation is made explicit. We discuss how, through interaction with their environment, gravitons may imprint this quantumness into some observable quantity. We then propose what this observable quantity might be.

\section{Perturbations during inflation} 

Let us briefly review the basics of quantum fluctuations generated during inflation. While this is textbook material, it is useful as to briefly setup the notation and make the article self-consistent for readers not specialized in the early Universe. More specialized readers can proceed directly to section~\ref{subsec:scalar-grav-grav}.

Any quantum field fluctuates due to the need to satisfy the equations of motion and the Heisenberg uncertainty principle, implemented through quantization of the field \cite{Mukhanov_2}. The simplest models of inflation include 2 fields: a scalar field $\phi$ called the \textit{inflaton} field, and the metric tensor field $g^{\mu\nu}$ that defines the spacetime. Let us first discuss the fluctuations of the inflaton field.

\subsection{Scalar perturbations}

We assume minimal coupling of the massive scalar inflaton field, with mass $m$,  $\phi(\mathbf{x})$ to gravity (in a spatially flat Friedmann universe). The action is \cite{Mukhanov_2}:
\begin{equation}
    \label{eq: action 1}
    S = \frac{1}{2}\int\sqrt{-g}\,g^{\mu\nu}\partial_\mu \phi \partial_\nu\phi + V(\phi) \,,
\end{equation}

where $V(\phi) = m \phi^2/2$. After substituting $g^{\mu \nu}=a^{-2}\eta^{\mu \nu}$ (where $\eta^{\mu\nu}$ refers to the Minkowski metric), $\sqrt{-g}=a^4$, and in terms of the auxiliary field $\chi(\eta,\mathbf{x})\equiv ~a(\eta)\phi(\mathbf{x})$, (\ref{eq: action 1}) becomes:
\begin{equation}
    \label{eq: action 2}
    S=\frac{1}{2}\int \text{d}^3\mathbf{x}\,\text{d}\eta\,\left\{\chi'^2-(\nabla \chi)^2 + \frac{a''}{a}\chi^2 + V(\phi)\right\}\,\,,
\end{equation}
where $'$ denotes derivative with respect to conformal time $\eta$, and $\nabla$ is a vector of spatial derivatives. The factor $a''/a$ is time-dependent; it accounts for the interaction of the scalar field with the expanding gravitational background. It implies that the energy of the scalar field is not conserved, which in quantum field theory leads to particle creation \cite{Mukhanov_2}. Since inflation is well approximated by de Sitter spacetime, 
\begin{equation}
    a(\eta)=-\frac{1}{H_\Lambda\eta}\,\,\,,\,\,\,-\infty<\eta<0
\end{equation}
the factor $a''/a$ becomes $2\eta^{-2}$. \\

We can expand the field $\chi(\eta, \mathbf{x})$ in its Fourier modes:
\begin{equation}
\label{eq: fourier expansion modes}
    \chi(\eta, \mathbf{x}) =\int\frac{d^3\mathbf{k}}{(2\pi)^{3/2}}\,\chi_\mathbf{k}(\eta)\,e^{i\mathbf{k}\cdot \mathbf{x}}\,\,.
\end{equation}
Varying the action \eqref{eq: action 2} with respect to $\chi$ and substituting the mode expansion \eqref{eq: fourier expansion modes}, we get the following differential equation for the scalar field:
\begin{equation}
\label{eq: DE for scalar field}
    \chi_k''+\left[k^2 + \frac{m^2}{H^2_\Lambda \eta^2}-\frac{2}{\eta^2}\right]\chi_k=0\,\,.
\end{equation}
The solutions can be written in general as:
\begin{equation}
    \label{eq: scalar Bunch Davis mode functions}
    \chi_\mathbf{k} = a_\mathbf{k} u_k(\eta) + a_{-\mathbf{k}}^\dagger u_k^*(\eta) \,\,,
\end{equation}
where $a_\mathbf{k}$ and $a_{-\mathbf{k}}^\dagger$ are integration constants, which upon quantization of the field will be promoted to annihilation and creation operators obeying the commutation relations:
\begin{equation}
    \label{eq: scalar comutation relations}
    \left[\Hat{a}_{\mathbf{k}},\,\Hat{a}_{\mathbf{k}'}^\dagger\right] = \delta(\mathbf{k}-\mathbf{k}')\,,\,\,\,\left[\Hat{a}_{\mathbf{k}},\,\Hat{a}_{\mathbf{k}'}\right] = \left[\Hat{a}_{\mathbf{k}}^\dagger,\,\Hat{a}_{\mathbf{k}'}^\dagger\right]=0\,\,.
\end{equation}

$u_k$ and $u_k^*$ are two linearly independent solutions to equation \eqref{eq: DE for scalar field}, which we call mode functions. They are normalized so $u_k {u_{-k}^*} ' - u_k^* u_{-k}'=i$. In the case of de Sitter spacetime , they are given in terms of Bessel functions (see \cite{Mukhanov_2}, pg. 89).\\

We can divide the behavior of a fluctuation of given wave number $k$ in its early and late time asymptotics.

At early times, $k|\eta|\gg 1$, the physical wavelength $\lambda_\text{ph}\sim a\,k^{-1}\simeq \frac{H_\Lambda^{-1}}{k|\eta|}$ is much smaller than the curvature scale of de Sitter $H_\Lambda^{-1}$, and the fluctuation behaves as in flat (Minkowski) spacetime. These are \textit{sub-horizon modes}. We can neglect the $\eta^{-2}$ term in \eqref{eq: DE for scalar field}, and choose the negative frequency mode to define the minimal excitation of the inflaton field (i.e. vacuum fluctuations) as \cite{Mukhanov_2}:
\begin{equation}
    u_k^{(\text{sub})}\approx \frac{1}{\sqrt{k}}e^{ik\eta}\,\,.
\end{equation}

As spacetime expands, $|\eta|$ decreases (remember $-\infty<\eta<0$), so for a given mode $k$, the physical wavelength is stretched until it becomes of order of the curvature scale $H_\Lambda^{-1}$ at time $\eta=\eta_k$, when $k|\eta|\sim 1$. Fluctuations of mode $k$ then cross the Hubble horizon and start to feel the curvature of spacetime. These are \textit{super-horizon modes}. At asymptotically late times, for $k|\eta|\ll 1$, the term $k^2$ in \eqref{eq: DE for scalar field} can be neglected, and we have solutions:
\begin{equation}
    \label{eq: super horizon modes}
    u_k^{(\text{super})} = A_k|\eta|^{2} + B_k|\eta|^{-1}\,\underset{\eta\rightarrow 0^-} {\longrightarrow} B_k|\eta|^{-1} \,\,\,\,\,\,(A_k,B_k \equiv \text{constant}).
\end{equation}\\

The amplitude of the fluctuations of the inflaton field is \cite{Mukhanov_2}:
\begin{equation}
    \label{eq: aplitude of fluctuations}
    \delta_\phi = \frac{1}{2\pi}a^{-1}k^\frac{3}{2}\,|u_k(\eta)| \approx \left\{\begin{array}{lr}
        \frac{1}{2\pi}\frac{k}{a} &;\,\,\lambda_\text{ph}\ll H_\Lambda^{-1} \\ & \\
       \frac{H_\Lambda}{2\pi}k^\frac{3}{2}  &  ;\,\, \lambda_\text{ph}\gg H_\Lambda^{-1}
    \end{array}\right.
\end{equation}
for sub- and super-horizon modes. We see, after horizon crossing at time $\eta_k\sim -k^{-1}$, the amplitude of the fluctuation remains constant in time; it \textit{freezes}. It will remain almost constant until inflation ends, when the physical size of the Hubble horizon will start growing (because of a decelerated expansion) and catch up with the fluctuation. The modes that became super-horizon the latest  (with smallest $k$) will be the first ones to reenter the horizon, and will be the earliest density perturbations of an otherwise homogeneous and isotropic Universe. 
The fluctuations that reentered right after the end of inflation had an effect on the smallest observable scales\footnote{Of course, the non-linear effects of gravity have completely disrupted these small scales at the present time.}, while those which are reentering our horizon now ($t=13.7$ Gyr) affect the largest scales, i.e. the large scale structure of our Universe.

The need of a gracefully exit of the inflationary stage requires that $H_\Lambda$ is not exactly constant, but decreases very slowly. In regard to the fluctuations, this will make those modes which left the horizon earlier have a slightly larger amplitude than the modes which left after (we say the spectrum is red-tilted towards larger scales). Note that we are not considering interactions terms that could modify the quantum-classical transition of the inflaton field \cite{Maldacena_2015,Campo:2007ar}. Indeed due to interaction terms, $k$ modes are not independent anymore and the environment could play an important role, for example \cite{Burgess_decoherence}.

\subsection{Tensor perturbations - gravitons}

We have seen how fluctuations of a massive scalar quantum field can indeed explain the primordial density inhomogeneities in the Universe. Lets now analyze the fluctuations of the other field, the metric tensor $g^{\mu\nu}$. The propagating modes corresponding to the \textit{transverse} and \textit{traceless} tensor fluctuations of the spacetime metric are what we call \textit{gravitons}. They behave as a minimally coupled, massless scalar field with two degrees of freedom (polarizations), and thus can be described by the same formalism used above for the scalar fluctuations (see \cite{Kolv_Turner}, chapter 8.4). This connection between the scalar and tensor sectors can be written
\begin{equation}
    h^{P}_{\mathbf{k}}(\eta) = \sqrt{16\pi G}\,\chi_{\mathbf{k}}(\eta)\,\,\,\,;\,\,\,\,(P=+,\times)
\end{equation}
where $\chi^{+,\times}_\mathbf{k}$ behave as two minimally coupled, real, massless scalar fields. $h_\mathbf{k}^P(\eta)$ are already the Fourier modes introduced in the expansion:
\begin{align}
    \label{eq: graviton def}
        h_{ij}(\eta, \mathbf{x})=\int\frac{d^3\mathbf{k}}{(2\pi)^{3/2}}\,\sum_{P=+,\times}\,\epsilon_{ij}^P(\mathbf{k})h_\mathbf{k}^P (\eta) \,e^{i\mathbf{k}\cdot\mathbf{x}}\,\,,
\end{align}
where $\epsilon^+ = \Hat{e}_x\otimes\Hat{e}_x - \Hat{e}_y\otimes\Hat{e}_y$ and $\epsilon^\times = \Hat{e}_x\otimes\Hat{e}_y + \Hat{e}_y\otimes\Hat{e}_x$ are the polarization tensors of the two graviton modes. This tensor polarization basis can be  expressed in matrix, by choosing $\Hat{\mathbf{z}}=\Hat{\mathbf{k}}$... as to visualize the matrix for these polarization tensors, form as:
\begin{equation}
    \epsilon^+(\Hat{\mathbf{z}})=\frac{1}{\sqrt{2}}\begin{pmatrix}
        1 & 0 & 0 \\
        0 & -1 & 0 \\
        0 & 0 & 0 
    \end{pmatrix}\,\,\,\,;\,\,\,\,\epsilon^\times(\Hat{\mathbf{z}})=\frac{1}{\sqrt{2}}\begin{pmatrix}
        0 & 1 & 0 \\
        1 & 0 & 0 \\
        0 & 0 & 0 
    \end{pmatrix}
\end{equation}

They have the following properties:
\begin{equation}
   \label{eq: polariz tensor orthogonality} \epsilon_{ij}^P(\mathbf{k})\epsilon_{ij}^{P'}(\mathbf{k}) = \delta^{PP'} \quad;\quad\quad
 \epsilon_{ij}^P(\mathbf{k})\epsilon_{ij}^{P'}(\mathbf{-k}) = s_P\,\delta^{PP'}\,\,\,,
\end{equation}
with $s_P=1$ for $P=+$ and $s_P=-1$ for $P=\times$. Note that we can have (at the same time or alternatively) two other different types of polarizations (L and R) that might be interesting for our purposes. In particular, it could create an intrinsic alignment that could be later measured as a signature of entanglement.

Quantization proceeds as in the scalar case; we expand the Fourier modes in their Bunch-Davies mode functions (same as for the scalar case) \cite{Mukhanov_2,Burgess_decoherence,Kanno_2019}: 
\begin{equation}
    \label{eq: Bunch Davies mode func for grav}
    h^{P}_\mathbf{k}(\eta)=\Hat{b}^{P}_\mathbf{k}u_{k}(\eta)+\left(\Hat{b}^{P}_{-\mathbf{k}}\right)^\dagger u_k^*(\eta)
\end{equation}
where $\Hat{b}_\mathbf{k}^P$ is the annihilation operator of a graviton with momentum $\mathbf{k}$ and polarization $P$ (idem for creation), with normalized commutation relations:
\begin{equation}
    \label{eq: commutation rel for grav}
    \left[\Hat{b}^P_{\mathbf{k}},\,\left(\Hat{b}^{P'}_{\mathbf{k}'}\right)^\dagger\right] = \delta(\mathbf{k}-\mathbf{k}')\delta_{PP'}\,,\,\,\,\left[\Hat{b}^P_{\mathbf{k}},\,\Hat{b}^{P'}_{\mathbf{k}'}\right] = \left[\left(\Hat{b}^P_{\mathbf{k}}\right)^\dagger,\,\left(\Hat{b}^{P'}_{\mathbf{k}'}\right)^\dagger\right]=0\,\,\,.
\end{equation}
Previous work on graviton entanglement was considered by \cite{Kanno:2017teu,Kanno_2019}

\subsection{Scalar-graviton-graviton interaction}
\label{subsec:scalar-grav-grav}
Here we present the interaction Hamiltonian for the scalar-graviton-graviton interaction, which will be needed later in section \ref{sec: Inflation as our lab}. We begin by expanding the action to 3rd order in perturbation theory \cite{Burgess_decoherence} in the Heisenberg representation and working on  the $\zeta$ gauge \cite{Maldacena:2002}:
\begin{equation}
    S_\text{int}^{(3)}=\frac{M_p^3}{8}\int dtd^3\mathbf{x}\,a\epsilon_1 \zeta \partial_l\gamma_{ij} \partial_l\gamma_{ij}
\end{equation}
which gives rise to the interaction hamiltonian (after $d\eta=dt/a$):
\begin{equation}
    H_\text{int}(\eta)=-\frac{M_p^3\epsilon_1 a^2}{8} \int d^3\mathbf{x}\,a\epsilon_1 \zeta(\eta,\mathbf{x}) \otimes \partial_l\gamma_{ij}(\eta,\mathbf{x}) \partial_l\gamma_{ij}(\eta,\mathbf{x})
\end{equation}
After changing $v\equiv aM_p\sqrt{2\epsilon_1}\zeta$ and $h_{ij} \equiv (1/2)aM_p\gamma_{ij}$, and taking the scale factor for the approximately de Sitter background $a\approx -(H\eta)^{-1}$:
\begin{equation}
    H_\text{int}^{(3)}\equiv G(\eta)\int d^3\mathbf{x}\, v(\eta,\mathbf{x})\otimes B_T(\eta,\mathbf{x})
\end{equation}
with
\begin{equation}
B_T(\eta,\mathbf{x}):=    \partial_l h_{ij}(\eta,\mathbf{x}) \partial_l h_{ij}(\eta,\mathbf{x})
\end{equation}
The function $G(\eta)=-\epsilon_1^{-1/2}\left[2\sqrt{2}M_pa(\eta)\right]^{-1}$ is derived in \cite{Burgess_decoherence}.\\

The tensor sector action is quadratic and thus represents free gravitons, it is of the type \eqref{eq: action 2}. The full interaction hamiltonian can be written:
\begin{equation}
\label{eq: full H graviton graviton inflaton}
    H_\text{int}^{(3)}= G(\eta)\int d^3\mathbf{x}\,\int \frac{d^3\mathbf{p}}{(2\pi)^{3/2}}\,v_\mathbf{p}(\eta)\,e^{i\mathbf{p}\cdot\mathbf{x}}\,\otimes \sum_{P=+,\times}\int \frac{d^3\mathbf{k}d^3\mathbf{q}}{(2\pi)^{3}} \,(\mathbf{k}\cdot\mathbf{q})\,\epsilon_{ij}^P(\mathbf{k})\epsilon_{ij}^{P'}(\mathbf{q})h_\mathbf{k }^P (\eta) \,h_\mathbf{q}^{P'} (\eta) e^{i(\mathbf{k}+\mathbf{q})\cdot\mathbf{x}}
\end{equation}\\
Substituting $v_\mathbf{p}(\eta)$ and $h_\mathbf{k,q}^{P,P'}$ by their expansions in Bunch-Davies mode functions, i.e. equations (\ref{eq: scalar Bunch Davis mode functions}) and (\ref{eq: Bunch Davies mode func for grav}), and using the commutation relations (\ref{eq: scalar comutation relations}) and (\ref{eq: commutation rel for grav}), we get, for the tensor sector:

\begin{align*}
    H_T^{(3)}\sim \sum_{P,P'=+,\times}& \int \frac{d^3\mathbf{k}\,d^3\mathbf{q}}{(2\pi)^{3}} \,(\mathbf{k}\cdot\mathbf{q})  \, e^{i(\mathbf{k}+\mathbf{q})\cdot\mathbf{x}}\,\epsilon_{ij}^P(\mathbf{k})\epsilon_{ij}^{P'}(\mathbf{q})\\
    &\left\{ \Hat{b}^{P}_\mathbf{k}\Hat{b}^{P'}_\mathbf{q} u_k u_q + s_{P'} \Hat{b}^{P}_\mathbf{k}\left(\Hat{b}^{P' }_{-\mathbf{q}}\right)^\dagger u_k u_q^* + s_{P} \left(\Hat{b}^{P}_{-\mathbf{k}}\right)^\dagger \Hat{b}^{P' }_{\mathbf{q}} u_k^* u_q + s_P s_{P'} \left(\Hat{b}^{P}_{-\mathbf{k}}\right)^\dagger \left(\Hat{b}^{P'}_{-\mathbf{q}}\right)^\dagger u_k^* u_q^*\right\},
    \label{eq: H_T general}
\end{align*}
where $H_T^{(3)}$ depends on $\mathbf{p}_\zeta$.

Now, this expression simplifies by taking $\mathbf{q}=-\mathbf{k}$, i.e., imposing that both gravitons are created with equal, opposite momenta. In the center of mass frame (CMF) of the massive inflaton, this implies, for modes inside the horizon, i.e. $|\mathbf{p}_\zeta| \sim \cal H(\eta)\ll |\mathbf{k}|$: 
\begin{equation}
\label{eq: CMF inflaton}
    \mathbf{p}_\zeta = 0 = \mathbf{k}+\mathbf{q}\,\,\,\Leftrightarrow\,\,\,\mathbf{q}=-\mathbf{k},
\end{equation}
where $\mathbf{p}_\zeta = 0$ in the CMF.
With this prescription, making use of the polarization tensor relations in \eqref{eq: polariz tensor orthogonality} and expanding the sum over polarizations explicitly:

\begin{align*}
    H_T^{(3)}(\mathbf{k}=-\mathbf{q}) & \sim  \sum_{P,P'=+,\times} \int \frac{d^3\mathbf{k}\,d^3\mathbf{q}}{(2\pi)^{3}} \,(\mathbf{k}\cdot\mathbf{q})  \, e^{i(\mathbf{k}+\mathbf{q})\cdot\mathbf{x}}\,\epsilon_{ij}^P(\mathbf{k})\epsilon_{ij}^{P'}(\mathbf{q})\, \left\{  \text{combination of }  b\,,b^\dagger \right\} \delta(\mathbf{k} + \mathbf{q})   \\
    &\sim \int\frac{d^3\mathbf{
    k}}{(2\pi)^{3}} \,k^2 \,e^{2i\mathbf{k}\cdot\mathbf{x}}
     \cdot \left\{ \left[  \Hat{b}^{+}_{\mathbf{k}} \Hat{b}^{+}_{\mathbf{k}} - \Hat{b}^{\times}_{\mathbf{k}} \Hat{b}^{\times}_{\mathbf{k}} \right] (u_k)^2  \right. + \\
    &  \quad \quad  \quad   \quad \quad \quad \quad \quad \quad + \left. \left[ \left(\Hat{b}^{+}_{\mathbf{k}}\right)^\dagger \Hat{b}^{+}_{\mathbf{k}}  + \left(\Hat{b}^{\times}_{\mathbf{k}}\right)^\dagger \Hat{b}^{\times}_{\mathbf{k}} + \left(\Hat{b}^{+}_{-\mathbf{k}}\right)^\dagger \Hat{b}^{+}_{-\mathbf{k}} + \left(\Hat{b}^{\times}_{-\mathbf{k}}\right)^\dagger \Hat{b}^{\times}_{-\mathbf{k}} \right] \left|u_k\right|^2 +  \right. & \\
    & \quad \quad \quad \quad   \quad \quad \quad \quad \quad   \left.  + \left[ \left(\Hat{b}^{+}_{-\mathbf{k}}\right)^\dagger \left(\Hat{b}^{+}_{\mathbf{k}}\right)^\dagger - \left(\Hat{b}^{\times}_{-\mathbf{k}}\right)^\dagger \left(\Hat{b}^{\times}_{\mathbf{k}}\right)^\dagger \right] \left(u_k^*\right)^2 \right\}
\end{align*}

The same procedure, done on the scalar sector of \eqref{eq: full H graviton graviton inflaton} yields (and imposing the momentum conservation mentioned above) where $H_{int}^{(3)}$ is $H_\zeta^{(3)} \otimes H_T^{(3)}$: 

\begin{align*}
    H_\zeta^{(3)} & \sim \int \frac{d^3\mathbf{p}}{(2\pi)^{3/2}}\,e^{i\mathbf{p}\cdot\mathbf{x}}\,v_\mathbf{p}(\eta) \\
    & \sim \int \frac{d^3\mathbf{p}}{(2\pi)^{3/2}}\,e^{i\mathbf{p}\cdot\mathbf{x}}\,\left( \Hat{a}_\mathbf{p}u_{p}+\Hat{a}_{-\mathbf{p}}^\dagger u_p^* \right)\,\underbrace{\delta(\mathbf{p}-\mathbf{k} - \mathbf{q})}_{=\delta(\mathbf{p})} \\
    & \sim \frac{1}{(2\pi)^{3/2}}\,\,\left( \Hat{a}_{\mathbf{0}}u_{0}+\Hat{a}_{\mathbf{0}}^\dagger u_{0}^* \right)  
\end{align*}

Besides factors, there is a term in the full hamiltonian (after imposing momentum conservation in the CMF of the inflaton, eq. \ref{eq: CMF inflaton}) that has 1 destruction operator for the inflaton, and 2 creation operators for the gravitons:
\begin{equation}
\label{eq: Entanglement hamiltonian}
    H_\text{int}^{(a\,b^\dagger b^\dagger)}\sim \int  \frac{d^3\mathbf{x}\,d^3\mathbf{k}}{(2\pi)^{3/2}} \,\frac{e^{2i\mathbf{k}\cdot\mathbf{x}}}{(2\pi)^{3}} \,k^2\,\Hat{a}_{\mathbf{0}} \otimes \left[ \left(\Hat{b}^{+}_{-\mathbf{k}}\right)^\dagger \left(\Hat{b}^{+}_{\mathbf{k}}\right)^\dagger - \left(\Hat{b}^{\times}_{-\mathbf{k}}\right)^\dagger \left(\Hat{b}^{\times}_{\mathbf{k}}\right)^\dagger \right] 
\end{equation}

\begin{figure}[H]
    \centering
    \includegraphics[width=0.4\textwidth]{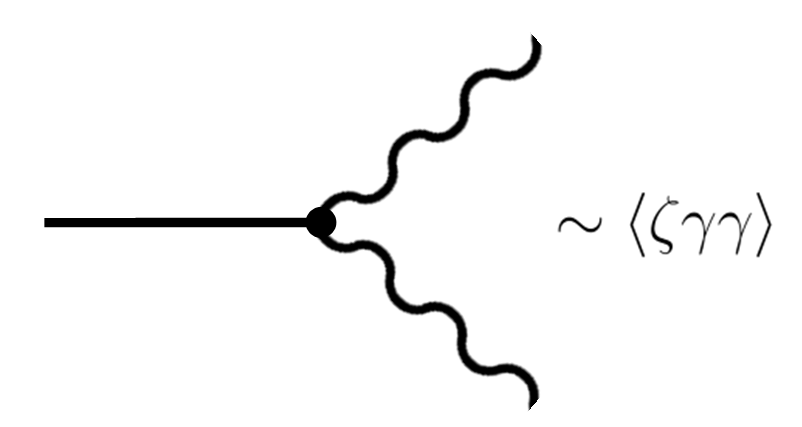}
    \caption{Three point vertex corresponding to inflaton-graviton-graviton (scalar-tensor-tensor) interaction. Gravitons are represented by wavy lines, the inflaton by the  straight line.}
    \label{fig:enter-label}
\end{figure}

There are several scenarios that will produce a classical coherent state for the inflaton following \cite{Gerry}. This can happen before the boost of for the case in which $|{\bf p}|\sim {\cal H}\ll |{\bf k}|\simeq |{\bf q}|$. This coherent state might be provided by a population of primordial Black Holes \cite{Piran_2023}, e.g see eq A.4 of \cite{Burgess_decoherence} taking a term like $\varphi J(t)$ where $J$ is a source of this coherent state of the inflaton. $J$ could be found, e.g., in the last line of eq. 3.8 of \cite{Maldacena:2002}. Another possibility is that the vacuum is not Bunch-Davies but a coherent state \cite{Polyakov:1982ug, Antoniadis:2006wq,Dvali:2017eba} (see also \cite{Kundu:2011sg, Ziaeepour:2015fqa}). More intuitively, all that is needed is a force in the inflaton potential, this could be given by a feature, or even the standard slow-roll. 

\section{Quantum nature, entanglement and Bell inequalities}

In this section, we briefly review what is known to be the ``most quantum aspect'' in quantum mechanics and quantum optics: \textbf{entanglement}. This effect makes the quantum nature of a system appear in a very explicit manner; if we are hoping to find traces of \textit{quantumness} in some observable, entanglement is what we should aim for.

\subsection{Definition of entanglement}
Entanglement between two quantum states is present when the quantum state describing the whole system cannot be written as a product of the states describing each subsystem separately, i.e. it is \textbf{non-separable}\footnote{Of course, this notion can be generalized to a system composed by $n\geq2$ single-particle states.}.\newline
For the case of two particles which can be either in one-particle states\footnote{These one-particle states can be thought of as being the corresponding eigenstates to the (real) eigenvalues of some hermitic operator $\Hat{O}$ corresponding to a physical observable $\mathcal{O}$.} $\ket{\psi_1}$ or $\ket{\psi_2}$, a general entangled state would the the superposed state:
\begin{equation}
\label{eq: generally entangled 2 state (no prescription)}    
    \ket{\Psi}=\alpha\ket{\psi_1}_1 \otimes \ket{\psi_1}_2+\beta\ket{\psi_2}_1 \otimes \ket{\psi_2}_2
\end{equation}
where $\alpha,\,\beta\in\mathbb{C}$, and $|\alpha|^2+|\beta|^2=1$. We note that $\ket{\psi_i}_j$ represents particle $j$ (the index $j$ could be a particle, momenta, directions, etc...]) in state $\ket{\psi_i}$.

Entanglement is exclusively a consequence of the principle of superposition present in quantum mechanics, and makes the non-locality of the theory explicit. Picture the following experiment:

\textit{One is to perform a measure of some observable $\mathcal{O}$ on the subsystem formed by particle 1 (for example, the same works if we start by measuring particle 2). This will result in the measure of one of the eigenvalues of the associated hermitic operator, for example $\lambda_1$, with probability $|\alpha|^2$, and the \textbf{instantaneous} collapse of the single-state of particle 1 to $\ket{\psi_1}$. If we look now at the 2-state describing the whole system, it will have instantaneously collapsed to its left term $\ket{\psi_1}_1\otimes\ket{\psi_2}_2$. In turn, this implies that from the moment particle 1 has been measured, particle 2 can only be in state $\ket{\psi_2}$, even if before the measurement on its companion, the state of particle 2 had contributions from both components $\psi_1$ and $\psi_2$. By performing a measure on 1 particle, we instantaneously change the state of the other.}\\

This effect is instantaneous upon measurement of particle 1 regardless of the positions of each of the particles, and thus non-local\footnote{The two particles can be in causally disconnected regions of spacetime and still be correlated at any instant in time through their entanglement.}. 

\subsection{Bell inequalities and experiment} \label{subsec: Bell ineq and experiment}

In 1964, the physicist John Stewart Bell proposed his famous theorem \cite{Bell}, which showed a clear mathematical difference between any description by a classical, local hidden variables theory, or by the quantum-mechanical theory that gave rise to the non-local effect of entanglement. Extensive work has been done over the years concerning all theoretical and experimental aspects of Bell's theorem and experiment \cite{Clauser_Shimony, CHSH, Gerry, Walls-Milburn}. For the purpose of our work, we are interested in replicating each of the needed elements of the Bell experiment in our inflationary paradigm. These elements are \cite{Maldacena_2015}:

\begin{itemize}
    \item Two separate spatial locations, call them Alice's location and Bob's location.
    \item An entangled state of the type \eqref{eq: generally entangled 2 state (no prescription)}, with components at these two locations.

    \item Two possible measurements of some physical observable, whose result is dependent on some local variable $\theta_i$ (or on some random choice between $\theta_i$ and $\theta_i'$), at each of the spatial locations (denoted by $i=1,2$). Each observation/measurement is represented by non-commuting operators, call them $A(\theta_1)$ and $A(\theta_1')$, and $B(\theta_2)$ and $B(\theta_2')$, respectively\footnote{This implies that upon two measurements of the same observable (one after the other), the result of the latter measurement is affected by the fact that the former has been previously measured; $[A(\theta_1),A(\theta_1')] \neq 0$ and $[B(\theta_2),B(\theta_2')] \neq 0$.}.
    
    \item Definite results for the quantum measurement of the operators. For the inequality presented below, both $A(\theta_1)$ and $B(\theta_2)$ measured on the state \eqref{eq: generally entangled 2 state (no prescription)} can yield $\pm 1$.
    
    \item A classical channel to transmit the results of the measurements to a common location where they can be correlated.
\end{itemize}

We can define the observable:

\begin{equation}
    S = C(\theta_1, \theta_2) + C(\theta_1', \theta_2) + C(\theta_1, \theta_2') - C(\theta_1', \theta_2')
\end{equation}

where $C(\theta_1, \theta_2)=\expval{A(\theta_1)\,B(\theta_2)}$. $S$ is such that in a description with a classical, local hidden variable theory (where the measure of $A$ is uncorrelated with the measure of $B$), $S^\text{local hidden variables}\leq 2$, while the quantum mechanical expectation value can be larger, $S^\text{QM}\leq 2\sqrt{2}$. By exceeding the former inequality in experimental measures, one proves the non-locality as an intrinsic property of entanglement and quantum mechanics. This particular Bell inequality is known as the Clauser-Horne-Shimony-Holt inequality \cite{CHSH}.

\begin{figure}
    \centering
    \includegraphics[width=0.8\textwidth]{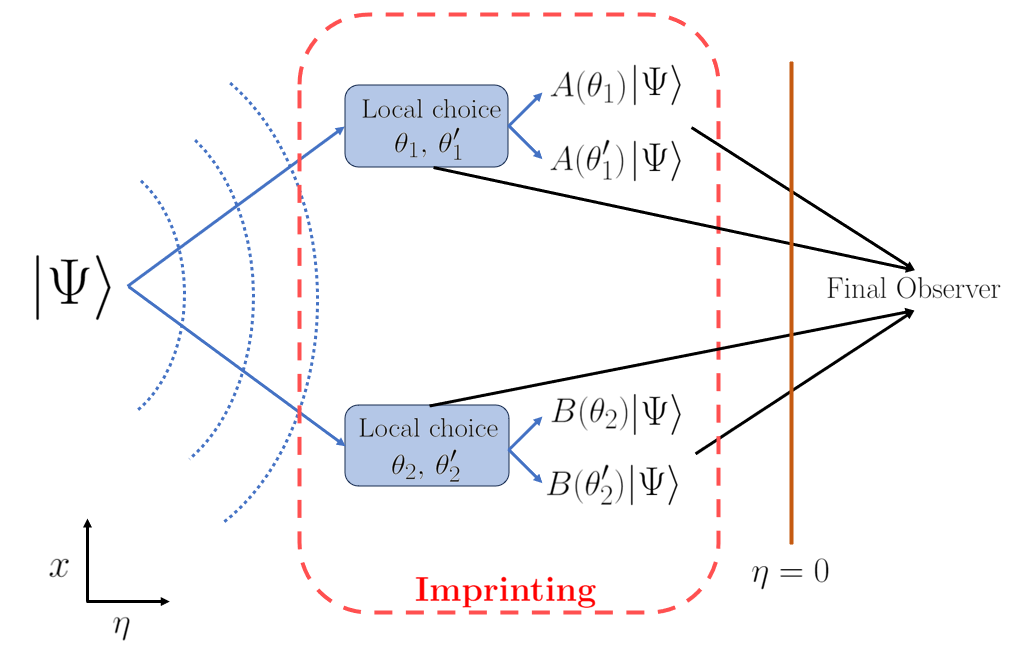}
    \caption{Set up for a Bell inequality violating experiment. The comoving time flows from left to right ($\eta=0$ represents the end of inflation), and spacial dimensions are drawn vertically. The entangled 2-state $\ket{\Psi}$ has components at two spatially-separate locations 1 (top) and 2 (bottom). Two different results of the measurement of observables $A$ and $B$ can be obtained, each dependent on the choice of the local variables $\theta_1$ and $\theta_2$, respectively. This choice is dependent on the interaction of the state with its local environment, it can be seen as a dependence on the measuring apparatus. The possible imprinting of the entanglement of state $\ket{\Psi}$ is represented by the dashed-line box. These results, together with the choices of $\theta_i$, are transmitted through classical channels (black arrows) to a final observer after inflation has finished.}
    \label{fig: maldacenas bell experiment setup}
\end{figure}

\subsection{Bell states}

As we have seen, an entangled 2-state is generally of the form \eqref{eq: generally entangled 2 state (no prescription)}. It is useful to work with 2-states known as \textit{Bell states} \cite{Gerry}, which are defined as
\begin{align}
    & \ket{\Psi_\text{Bell}^\pm}=\frac{1}{\sqrt{2}}(\ket{0}_1\ket{1}_2\pm \ket{1}_1\ket{0}_2)\,, \label{eq: bell state psi +-} \\
    & \ket{\Phi_\text{Bell}^\pm}=\frac{1}{\sqrt{2}}(\ket{0}_1\ket{0}_2\pm \ket{1}_1\ket{1}_2)\,. \label{eq: bell state phi +-}
\end{align}
Here, $\ket{0}_i$ and $\ket{1}_i$ represent the basis states of a general bipartite system\footnote{A bipartite or two-mode system is a single-particle system with only two non-degenerate eigenvalues for a given observable $\mathcal{O}$, such that the one-state can be written as $\ket{\psi}_i^\text{1-state}=a\ket{0}_i + b\ket{1}_i\,\,$, with $a,b\in\mathbb{C}$, and $|a|^2 + |b|^2 = 1$.} (or two-mode system). The notation $\ket{i}_1\otimes \ket{j}_2$ denotes the tensor product of the basis states of subsystems 1 and 2, which define a basis of the Hilbert space $\mathcal{H}=\mathcal{H}_1\otimes\mathcal{H}_2$, in which two-particle states live. 

\section{Inflation as our laboratory} \label{sec: Inflation as our lab}

We have now all the necessary ingredients to design a plausible cosmological Bell-type experiment taking place during inflation. Some signature of violation of Bell inequalities, and therefore of the non-locality intrinsic to the quantum theory would allow us to prove the quantum mechanical origin of primordial in-homogeneities that gave rise to the observed large scale structure of our Universe. At least, this possible signature might serve as prove of concept for the theoretical scientific community, and serve as a possible observational path to follow by the experimental community.\\

\subsection{Elements of the cosmological Bell experiment}

We proceed to describe our realization of a cosmological Bell-type experiment in the inflationary epoch. For that we need a specific realization of the general elements described in section \ref{subsec: Bell ineq and experiment}, and goes as follows:

\begin{itemize}
    \item Alice's and Bob's locations will be two spatially separate points at the inflationary Hubble horizon, $H^{-1}_\Lambda\approx \text{constant}$, i.e., when the measurement of the polarization of the gravitons is performed, the momentum of the gravitons (when they are produced) should be around or greater than $ \cal H$ and the momentum of Alice and Bob (i.e. related to the inverse of the dimensions of their "laboratories") should be much greater than that of the graviton. 
    
    \item Our Bell state describes two gravitons entangled in their tensor polarizations, see section \ref{subsec: entangled state of gravitons}, eq. \eqref{eq: grav entangled in polariz}. Each graviton is at one of the spatial locations (Alice's graviton and Bob's graviton).

    \item The physical observable is the polarization of the gravitons, with two possible results $\theta_i\equiv P_i = +, \times$ (we could use also L and R polarizations depending on how Alice and Bob do the measurements.) The measure at each location is performed by the decoherence effect due to the gathering of ``which-path information'' by the cosmological horizon as described in \cite{Wald_BH_decohere, Wald_Killing_horiz_decohere}, see section \ref{subsec: measure by cosmo horiz}. 
    
    \item The possible results of the measurement (decoherence) of Alice's and Bob's gravitons are the $\times$ and $+$ polarizations of each graviton. In our case, the result would then leave an imprint through the interaction of two inflatons with the polarized local curvature (generated by the graviton). At each location Alice and Bob, we have scattering of two inflatons through graviton exchange (GE), as described in section \ref{subsec: Imprint through GE} (see also \cite{Burgess_decoherence} for a discussion on GE). We could say that both Alice and Bob use a pair of inflatons each to imprint the results of their measure of polarization.

    \item The result of the measure (or its imprint) is classically transmitted by the propagation of these pairs of scalar fluctuations (inflatons) after horizon crossing, where their amplitude freezes until the end of inflation, and then reenters during radiation epoch. Note that we have a non-trivial effect due to the derivatives on the two scalar fluctuations (se eq. 5.1 in \cite{Burgess_decoherence}) this provides a non-trivial fingerprint that depends on the polarization of the graviton that Alice and/or Bob measured in their patch. The common location to which the results are transmitted is the range of scales set between the corresponding moments of reentering of Alice's fluctuations and Bob's fluctuations. Results can be correlated through the 4-point functions $\expval{\zeta_{\mathbf{k}_1} \zeta_{\mathbf{k}_2} \zeta_{\mathbf{k}_3} \zeta_{\mathbf{k}_4}}^\text{GE}_\mathbf{A}(P_1)$, $\expval{\zeta_{\mathbf{k}_1} \zeta_{\mathbf{k}_2} \zeta_{\mathbf{k}_3} \zeta_{\mathbf{k}_4}}^\text{GE}_\mathbf{B}(P_2)$ 
\end{itemize}
The derivatives on the scalar field fluctuations could leave an additional imprint on the two subhalos of each patch with the possibility that we could observe it on the large scale structure through the 'intrinsic alignment'. Note that this 'intrinsic alignment' is obtained by performing the (perturbed Taylor) expansion between one subhalo with the second one in the same patch. In this case, the two derivatives are obtained for the same potential: i.e. a tidal effect between these two subhaloes linked to the polarization of the graviton.

In the following, we present a feasible way in which each of these elements appear in our inflationary setup.

\subsection{Entangled state of gravitons} \label{subsec: entangled state of gravitons}

We start with the Hamiltonian \eqref{eq: Entanglement hamiltonian}.
We now label the gravitons with momentum $-\mathbf{k}$ and $\mathbf{k}$ as gravitons 1 and 2, respectively. Following \cite{Gerry}, and ignoring the left hand side of the tensor product in the Hamiltonian (pumping signal, corresponds to the destruction operator of the coherent inflaton field) \\

\begin{equation}
    H_\text{int}^{(a\,b^\dagger b^\dagger)} \equiv H_I \sim \left[ \left(\Hat{b}^{+}_{1}\right)^\dagger \left(\Hat{b}^{+}_{2}\right)^\dagger - \left(\Hat{b}^{\times}_{1}\right)^\dagger \left(\Hat{b}^{\times}_{2}\right)^\dagger \right] + h.c.
\end{equation}

We start from an initial vacuum state:
\begin{equation}
    \ket{\Psi_0}= \ket{0}\equiv\ket{0}_1\otimes\ket{0}_2 \equiv \ket{0}_{+,1}\ket{0}_{\times,1}\otimes\ket{0}_{+,2}\ket{0}_{\times,2}
\end{equation}

where we have split the two vacua (for gravitons 1 and 2) in their ``plus'' and `cross'' polarization components, i.e. we define the single states of each graviton (and its vacuum) in terms of these two components:

\begin{align*}
    \label{eq: def vacum, plus,cross} 
    \ket{0}_i := \ket{0}_{+,i}\ket{0}_{\times,i} \quad ; \quad
    \ket{+}_i:=\ket{1}_{+,i}\ket{0}_{\times,i}\quad ; \quad
    \ket{\times}_i := \ket{0}_{+,i}\ket{1}_{\times,i}   \quad ; \quad (i=1,2)
\end{align*}
We will discuss below the values for the momenta of the gravitons.
Taking a first order approximation in the Schrödinger equation for the time evolution of the state:
\begin{equation}
    \ket{\Psi (t)}=e^{-itH_I/\hbar}\ket{\Psi_0}\approx \left(1-\frac{it}{\hbar} H_I \right) \ket{0}\,\,\,,
\end{equation}
where we expand the evolution operator to first order because: 1) we are considering instantaneous interactions and 2) to obtain the Bell state we consider very short inflaton interactions (in the coherent (classical) state) and the two gravitons. This leads to

\begin{equation}
\label{eq: grav entangled in polariz}
    \begin{aligned}
    \ket{\Psi (t)} & \sim  \ket{0} - \frac{it}{\hbar}\left(\ket{1}_{+,1}\ket{0}_{\times,1} \otimes \ket{1}_{+,2}\ket{0}_{\times,2} - \ket{0}_{+,1}\ket{1}_{\times,1} \otimes \ket{0}_{+,2}\ket{1}_{\times,2}\right)  \\
     & = \ket{0} - \frac{it}{\hbar} \left(\ket{+}_1\ket{+}_2 - \ket{\times}_1\ket{\times}_2\right) \\
     &\equiv \frac{it}{\hbar}\left( \ket{\times\times} - \ket{++} \right)
\end{aligned}
\end{equation}

We can see that this state $\ket{\Psi}$ is proportional to the Bell state $\ket{\Phi^-_\text{Bell}}$ in \eqref{eq: bell state psi +-}.

\subsection{The measure by the cosmological horizon} \label{subsec: measure by cosmo horiz}

Once we have our entangled state of graviton polarizations $\ket{\Psi}$ of equation \eqref{eq: grav entangled in polariz}, we can proceed with the measurement of the state as part of the Bell experiment. The two gravitons need to be spatially separated. This is reasonable since they come from the inflaton ``pump'' (which we set in the CMF of $\zeta$ and got gravitons with $-\mathbf{k}$ and $\mathbf{k}$), and we can always boost them to a lab frame in which they are at an angle $2\alpha = \arccos\left[\left(E_\zeta/\sqrt{2}\,E_\gamma\right)^2 - 1\right]$ at the moment of creation. Here $E_\zeta$ and $E_\gamma = k$ are the energies of the inflaton and gravitons, respectively. Since teh boost is small for our reference frame, the change in energy is also negligible. 

Now, in order to perform the measurement of our Bell state, we use the results in \cite{Wald_BH_decohere,Wald_Killing_horiz_decohere}. In these works, it is suggested that in the presence of a Killing horizon, information from a quantum state in the vicinity of the horizon falls through it, and this decoheres the quantum superposition (or measures it)\footnote{A measurement performed on a quantum state collapses the state to one of the eigenstates of the measured operator. This destroys the state, just as decoherence does.}. This Killing horizon can come from different objects, including the \textit{cosmological horizon} intrinsic to de Sitter spacetime \cite{Wald_Killing_horiz_decohere}. It could also be due to the presence of a population of primordial black holes during inflation \cite{Piran_2023}; also note that we are not in an exact de Sitter spacetime and that this symmetry (i.e. the existence of this Killing horizon) could be weakly broken.

This result comes from the fact that a massive object in a quantum superposition radiates a retarded gravitational field (soft gravitons). In Minkowski spacetime, for an inertial frame and imposing adiabatic evolution of the quantum state, decoherence can be avoided (a recombination of the spatially separated wave-packets of the superposed state can be performed in a way such that no information is lost, see \cite{Wald_BH_decohere}). However, in the presence of a Killing horizon like the cosmological horizon of inflation, part of this radiated field inevitably falls through the horizon. In this way, as time passes the horizon gathers ``which-path information'' about the quantum superposition, decohering (measuring) more and more the state. This decoherence is expressed as:
\begin{equation}
    \label{eq: amount of decoherence}
    \mathcal{D} = 1 - \braket{\psi_1}{\psi_2}_{H_\Lambda^{-1}} = 1 - e^{-\frac{1}{2}\expval{N}}
\end{equation}
where $\expval{N}$ represents the average number of soft gravitons (gravitational field) that are radiated through the
horizon and $\ket{\psi_1},\,\ket{\psi_2}$ represent the quantum states of the radiated gravitational field sourced by the object in a spatial quantum superposition, in our case each graviton at locations $A$ and $B$\footnote{The notation in Eq. \ref{eq: spatial separated wavepackets} is used to stress the fact that each graviton is at a different spatial location A and B. We stress that this notation does not represent that the quantum 2-state is separable, since we have an entangled state. This is why we write $\left\{\times\text{ or } +\right\}$ as being the two polarization eigenstates for each spatially separated graviton.}:
\begin{equation}
\label{eq: spatial separated wavepackets}
    \ket{\Psi} \sim \frac{1}{\sqrt{2}}\left(\ket{\{\times \,\text{or}\,+\}}_A + \ket{\{\times \,\text{or}\, +\}}_B\right) \equiv \frac{1}{\sqrt{2}}\left( \ket{\psi_1} + \ket{\psi_2}\right)
\end{equation}
In Eq. \eqref{eq: amount of decoherence} $\braket{\psi_1}{\psi_2}_{H_\Lambda^{-1}}$ is the inner product (defined in \cite{Wald_BH_decohere}) of the 1-particle states of the soft gravitons falling through the horizon $H_\Lambda^{-1}$. Note that the massive quantum particle that is subjected to decoherence has ``somewhere" imprinted the information of the polarization of the graviton discussed previously.

One issue that arises is the fact that gravitons are massless. We will deal with this in sec. \ref{subsec: Imprint through GE}. For now, we take that decoherence occurs through the gathering of ``which-path information'' by the cosmological horizon. Let us detail how exactly decoherence happens. There are two options, both leading to an ``effective measurement'' of the polarization:

\begin{enumerate}
\item The entangled 2-state decoheres as a whole to only one of the superposed states, e.g. 
$$\ket{\Psi} \sim \ket{\times\times} - \ket{++} \overset{\mathcal{D}}{\longrightarrow}\ket{\times \times}\,\,,$$ which is something desirable, because it collapses our polarizations to one possibility only.\\

\item The decoherence occurs separately on the superposed states of one graviton (or both) (e.g. Alice's graviton): $$\ket{\psi}_A\sim\ket{\{\times\text{ or } +\}}_A\overset{\mathcal{D}_{A}}{\longrightarrow} \ket{\times}_A\,\,,$$ which, if (some) coherence of the entangled 2-state is maintained, would automatically force Bob's graviton to $\ket{\times}_B$.\\

\end{enumerate}

The right hand term in the second equality of Eq. \eqref{eq: amount of decoherence} contains $\expval{N}$, which represents the average number of soft gravitons (gravitational field) that are radiated through the horizon. In the case of a cosmological horizon, we have \cite{Wald_Killing_horiz_decohere}:
\begin{equation}
    \expval{N}\sim \frac{m^2 d^4}{R_H^5}\,T\,\,,
\end{equation}
where $R_H = H_\Lambda^{-1}$ is the cosmological horizon radius, $m$ is the mass of the superposed quantum object (see sec. \ref{subsec: Imprint through GE}), $d$ is the spatial separation between the two wavepackets Alice and Bob, and $T$ is the proper time for which the mass has been radiating soft gravitons that fall through the horizon. Note that the lab A is at the worldline $r=0$ (inertial orbit of the static Killing field), being the horizon at $r=R_H$. Alice and Bob should be in opposite directions at a distance given by \ref{eq: distance wavepackets}. This is true for our gravitons with comoving wavenumber $k$, since they where created at the same spatial location which we take it to be $r=0$, and in comoving coordinates are in an inertial frame. 

This leads to a decoherence time (in natural units):
\begin{equation}
\label{eq: decoherence time}
T_D\sim \frac{R_H^5}{m^2 d^4}   \,\,.
\end{equation}

The proper distance $d$ around the time of horizon crossing can be estimated as:
\begin{equation}
\label{eq: distance wavepackets}
    d \approx 2R_H\sin\alpha \approx 2R_H \sin\left(\frac{1}{2}\,\arccos\left[\left({E_\zeta}/{\sqrt{2}\,k}\right)^2 - 1\right]\right)\,\,.
\end{equation}

Now, under the decoherence/measurement induced by the cosmological horizon, the polarization of \textit{both} gravitons will be (partially\footnote{Since decoherence might only partially occur, i.e. $0 < \mathcal{D} < 1$, a full collapse to one of the polarization eigenstates might not occur, but only a partial tendency to it.}) forced to the same result of the measure, either $\times$ or $+$ polarization, right before horizon crossing.

\subsection{Imprinting - scattering of inflatons by polarized graviton exchange at horizon crossing}  \label{subsec: Imprint through GE}

Let us now hint a possible way in which the measure of the entangled state of gravitons\footnote{A full quantitative calculation, with cosmological observables, will be presented in a future publication.} might leave an imprint on a physical observable after inflation. This imprinting should happen right before horizon crossing, so that after becoming super-horizon the amplitude of fluctuations freezes and is classically preserved until its reentering into our Hubble horizon.\\

As we saw, we can use decoherence by the cosmological horizon as a measuring apparatus, as long as we have some mass associated to each graviton. We suggest that this effective mass comes from the Lagrangian and will enable the imprinting of the non-locality of the entangled state.\\

\begin{figure}
    \centering
    \includegraphics[width = 0.6\textwidth]{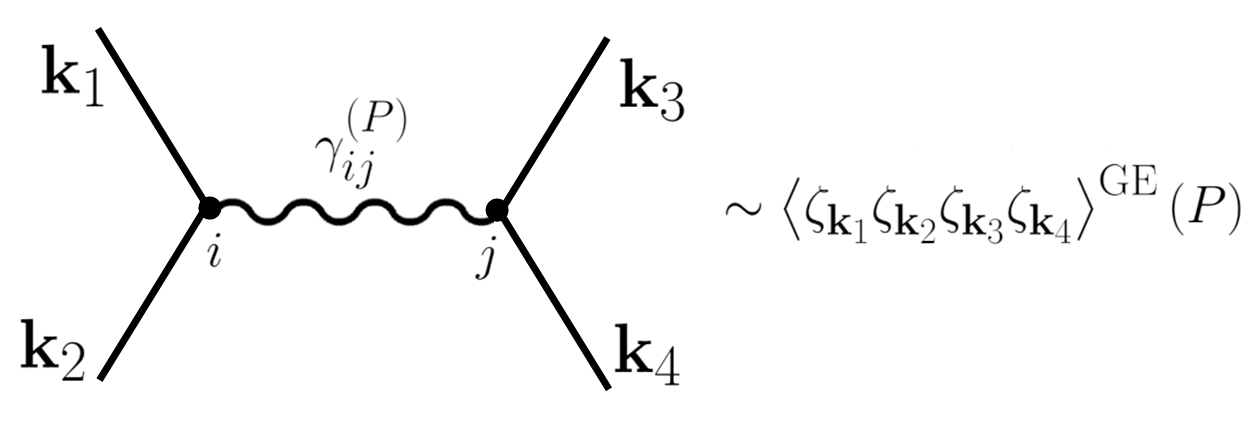}
    \caption{Graviton exchange of four inflatons \cite{Bellomo_2018}. Process of measure-imprinting at horizon crossing. The graviton is entangled with its partner, as described in the text. Note that the graviton on the diagram is in fact the two soft gravitons.}
    \label{fig: feynmann diag for 4 scalar int}
\end{figure}

\begin{figure}
    \centering
    \includegraphics[width=0.95\textwidth]{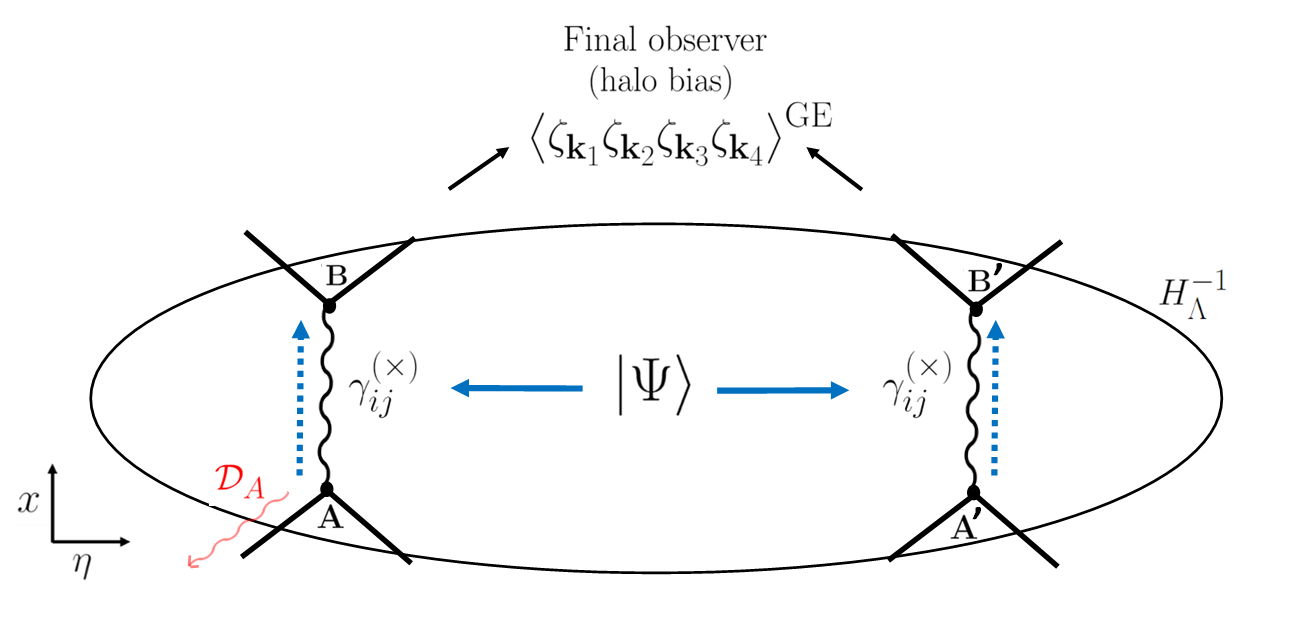}
    \caption{Schema of the imprinting process described in Sec. \ref{subsec: measure by cosmo horiz} and \ref{subsec: Imprint through GE}. The entangled state (in graviton polarizations) $\ket{\Psi}$ is transmitted to two separate spatial locations. Around the time of horizon crossing, two inflatons arrive and give some effective mass to each graviton while undergoing scattering by GE. This effective mass produces the radiation of soft gravitons (represented by the red, wavy line) through the horizon $H_\Lambda^{-1}$, and makes the entangled state decohere. This is represented for the case of graviton A, by $\mathcal{D}_A$, which collapses to $\times$ polarization. The blue, dotted arrow represents the instantaneous effect of collapse to $\times$ polarization of graviton $B$ due to the entanglement. This effect then remains on the 4-point function $\expval{\zeta_{\mathbf{k}_1}\zeta_{\mathbf{k}_2} \zeta_{\mathbf{k}_3} \zeta_{\mathbf{k}_4}}^\text{GE}_\mathbf{A,B}$. The fluctuations freeze after horizon crossing, and classically transmit this ``spatial preference'' (from polarization preference) to a final ``observer'' after the end of inflation ($\eta=0$). In our case, this final observer could be the correlation between 4-point functions $A$ and $B$ in the halo bias (calculated in \cite{Bellomo_2018}).}
    \label{fig: GE imprinting}
\end{figure}

Take the four-point correlation function between scalar fluctuations (inflatons) mediated by graviton exchange (GE), depicted in Fig. \ref{fig: GE imprinting}. As shown in \cite{Bellomo_2018} the graviton exchange can be manifested as a measurable signal in the halo bias. In the calculation of the scattering amplitude of the shown Feynman diagram of Fig. \ref{fig: feynmann diag for 4 scalar int}, performed in \cite{Seery_2009_Trispec_GE} and used in \cite{Bellomo_2018}, the sum over all possible polarizations of the graviton is taken, as is standard in QFT calculations when the polarization of the exchanged particle (graviton in this case) is unknown. The polarization of the graviton taking part in the GE is known and contained in the fig.~\ref{fig: feynmann diag for 4 scalar int} and therefore the overall scattering amplitude might be different. This then could have an effect on $\expval{\zeta_{\mathbf{k}_1} \zeta_{\mathbf{k}_2} \zeta_{\mathbf{k}_3} \zeta_{\mathbf{k}_4}}^\text{GE}(P)$, where $P$ represents the polarization of the exchanged graviton. We can think of this scattering as two inflatons which, at the time right before horizon crossing, feel some local curvature generated by the mediating graviton. Since the tensor polarizations correspond to the directions in which these local spacetime oscillations occur, $+$ and $\times$, one can picture a qualitative difference when the scattered inflatons ``feel'' this oscillatory feature. \\

We suggest that it is the mass of the incoming inflatons right around the time of the graviton exchange that gives an effective mass to the graviton so that it can radiate soft gravitons that fall through the horizon $H_\Lambda^{-1}$, and make it decohere through the mechanism presented in the previous section. For decoherence to become effective \textit{while} the GE occurs, the time of decoherence should be comparable with the characteristic time scale of gravitation, for example if 
\begin{equation}
\label{eq: GE time = T_D}
    T_D \sim t_\text{GE} \sim\frac{1}{M_\text{Pl}}\,\,.
\end{equation}

From Eqs. (\ref{eq: decoherence time}, \ref{eq: distance wavepackets}) and \ref{eq: GE time = T_D}, taking comoving distance in comoving coordinates $\Tilde{d} = d/a(\eta)$  and conformal time $\eta = T/a(\eta)$, and substituting $a(\eta)= R_H/\eta$ at the time around horizon crossing $|\eta|\sim k^{-1}$ for a mode of fixed comoving wavenumber $k$, we obtain the following relation (besides order one factors):
\begin{equation}
    \label{eq: condition for decoherence of mode k}
    k_D\sim \left(\frac{m^2}{M_\text{Pl}\,R_H^5}\right)^{1/4} \sim \left(\frac{m^2}{M_\text{Pl}}\right)^{1/4}\,H_\Lambda^{5/4}
\end{equation}
As mentioned, the effective mass of the decohering graviton would be twice the mass of the incoming inflatons, i.e. $m\equiv m_\text{eff}^{(\gamma)} \approx m^{(\zeta)}_1 + m^{(\zeta)}_2$. $k_D$ represents the modes that upon arrival of the two inflatons, would decohere right before horizon crossing. Taking $m\sim10^{-3} M_\text{Pl}$ and $H_\Lambda\sim\varepsilon_\Lambda^{1/2}$, yields:
\begin{equation}
    k\sim 10^{-3/2} \varepsilon_\Lambda^{5/8} \ll 1
\end{equation}
This number is expected to be small, and thus the modes that decohere while the GE takes place and could imprint the gravitons polarization are those corresponding to large scales. Also note that the graviton could decohere also with Hawking radiation.\\

Let as now put together this possible imprinting with the fact that the graviton taking part in the GE has an entangled partner\footnote{Self coupling of the inflaton is negligible compared to GE (enters in four point function $\expval{\zeta_{\mathbf{k}_1} \zeta_{\mathbf{k}_2} \zeta_{\mathbf{k}_3} \zeta_{\mathbf{k}_4}}$).}. 
Concretely, imagine that Alice's graviton decoheres because of the which-path information obtained by the cosmological  horizon i.e. $\mathcal{D}_A>0$, and collapses to ``cross'' polarization, $\times$. Now, only the ``cross'' term will contribute to the GE correlation through Alice's graviton, $\expval{\zeta_{\mathbf{k}_1} \zeta_{\mathbf{k}_2} \zeta_{\mathbf{k}_3} \zeta_{\mathbf{k}_4}}^\text{GE}_\mathbf{A}(\times)$. And more importantly, from the moment $\mathcal{D}_A$ has become effective, some other scattering of two new inflatons through GE of Bob's graviton will undergo exactly the same effect, i.e. only $\expval{\zeta_{\mathbf{k}_1} \zeta_{\mathbf{k}_2} \zeta_{\mathbf{k}_3} \zeta_{\mathbf{k}_4}}^\text{GE}_\mathbf{B}(\times)$ will contribute. In this way, because of the fact that there was an entangled 2-state between gravitons A and B, a spatial preference could have been formed. This preference would not have been there if our theory had preserved locality, or if we did not have an entangled state. This is shown in Fig.~\ref{fig: GE imprinting}.\\

Let us interpret this process as follows: If we think that the ``graviton" in Fig.~\ref{fig: feynmann diag for 4 scalar int} is, in reality our two gravitons that are emitted in two different directions to lab Alice and lab Bob, then the two Feymann diagrams are in reality the two measures that we need to do for each lab, i.e.,  where we will do a measure and they will be in the two sub-halos of patch A(lice) and B(ob).

\section{Conclusions}
We have presented a mechanism during the inflation period to generate entangled states of gravitons using the inflaton field as a ``pump''. This entangled states can then be measured at a killing horizon, thus resulting in a observational feature of quantumness in the early Universe. Note that our effect is due to second order effects; we have a non-trivial effect due to the derivatives on the two scalar fluctuations and this provides a fingerprint that depends on the polarization of the graviton that Alice and/or
Bob measured in their patch.
We have proposed possible signatures in the halo bias due to graviton exchange, but it is also possible to speculate that this signature could manifest in the intrinsic alignment of galaxies or influence the spin of black holes if a primordial population exists during inflation. We will quantify the observational signatures in a future publication. 

\section*{Acknowledgments}

Funding for the work of PT and RJ was partially provided by project PGC2018-098866- B-I00 y
FEDER “Una manera de hacer Europa”, and the “Center of Excellence Maria de Maeztu
2020-2023” award to the ICCUB (CEX2019- 000918-M) funded by \\
MCIN/AEI/10.13039/501100011033
 DB acknowledges partial financial support from the COSMOS
network (www.cosmosnet.it) through the ASI
(Italian Space Agency) Grants 2016-24-H.0, 2016-24-H.1-2018 and
2020-9-HH.

\bibliographystyle{unsrt}


\end{document}